\documentclass{article}
\usepackage{ijcai20}

\usepackage{times}

\usepackage{soul}
\usepackage{url}
\usepackage[utf8]{inputenc}
\usepackage[small]{caption}
\usepackage{subcaption}
\usepackage{graphicx}
\usepackage{amsmath,amssymb,amsthm}
\usepackage{booktabs}
\usepackage{xcolor}
\usepackage[ruled,vlined, linesnumbered, boxed]{algorithm2e}
\usepackage{tikz}
\usepackage{pgfplots}
\pgfplotsset{compat=1.14}
\usepackage{makecell}
\usepackage{bm}
\usepackage{thmtools}
\usepackage{thm-restate}
\usepackage[hang,flushmargin]{footmisc} 
\usepackage{expl3}

\title{Game Theory on the Ground: \\The Effect of Increased Patrols on Deterring Poachers}

\author{
 Lily Xu$^1$\footnote{Contact Author}\and
 Andrew Perrault$^1$\and
 Andrew Plumptre$^2$\and
Margaret Driciru$^3$\and\\
Fred Wanyama$^3$\and
Aggrey Rwetsiba$^3$\And
Milind Tambe$^1$
 \affiliations
 $^1$Harvard University, 
 $^2$Key Biodiversity Secretariat, 
 $^3$Uganda Wildlife Authority
 \emails
 \{lily\_xu, aperrault\}@g.harvard.edu,
 aplumptre@keybiodiversityareas.org,
\{margaret.driciru, fred.wanyama, aggrey.rwetsiba\}@ugandawildlife.org,
milind\_tambe@harvard.edu,
}

\begin{document}

\maketitle

\begin{abstract}
Applications of artificial intelligence for wildlife protection have focused on learning models of poacher behavior based on historical patterns. However, poachers' behaviors are described not only by their historical preferences, but also their reaction to ranger patrols. Past work applying machine learning and game theory to combat poaching have hypothesized that ranger patrols deter poachers, but have been unable to find evidence to identify how or even if deterrence occurs. Here for the first time, we demonstrate a measurable deterrence effect on real-world poaching data. We show that increased patrols in one region deter poaching in the next timestep, but poachers then move to neighboring regions. Our findings offer guidance on how adversaries should be modeled in realistic game-theoretic settings. 
\end{abstract}

\maketitle

\section{Introduction}

Illegal wildlife poaching threatens scores of endangered animals, from elephants and tigers to turtles and seahorses. 
The majority of funds---over 1.3 billion USD each year---invested to combat the illegal wildlife trade goes towards protected area management \cite{gulland2019illegal}. Unfortunately, the majority of protected areas are still under-resourced, with too few rangers to patrol these vast lands. Improving the efficacy of ranger patrols is therefore imperative to protecting wildlife. 


Artificial intelligence (AI) has been leveraged to prevent poaching, focusing on learning poacher behavior to plan ranger patrols \cite{gholami2018adversary}. During these patrols, rangers directly protect wildlife by confiscating snares that would otherwise sit out and trap endangered animals. An indirect impact of patrols to protect wildlife is through \textit{deterrence}, reducing the frequency at which poachers attack in the future. 
Past work has investigated deterrence to inconclusive results \cite{ford2017real,dancer2019evaluation}. Previous models typically accounted for deterrence simply by including past patrol effort as a feature in machine learning models, without specifying any behavior of the effect of past patrol effort \cite{xu2020stay}. 


Deterrence is believed to be the dominant means by which patrols reduce illegal activity \cite{levitt1998increased}, as rangers rarely apprehend poachers and only remove an estimated 10\% of snares \cite{moore2018ranger}. To justify the high cost of ranger patrols, we should therefore expect a significant deterrence effect. 
However, the challenge of demonstrating deterrence in real-world poaching data remains unanswered. 




In this paper, for the first time, we demonstrate measurable deterrence, where increased patrolling decreases the likelihood of poaching in future timesteps, and also \textit{displacement} in that poachers move toward nearby regions. Using real-world poaching data from Queen Elizabeth National Park (QENP) and Murchison Falls National Park (MFNP) in Uganda, we study the effect of varying levels of patrol effort on poacher response. 

We find that (a)~deterrence can be measured at the one square kilometer resolution; (b)~the amount of patrol effort in kilometers patrolled determines the strength of the deterrence effect, not simply whether or not a target is visited; (c)~ranger observations of illegal activity has the greatest impact, not simply past patrol effort; and (d)~increased patrols in one region cause poachers to become more active in nearby regions. 
Our findings help guide future research, both in predictive modeling and game theory. Specifically, we suggest that an accurate adversarial behavioral model should account for past patrol effort in both the individual region and also neighboring regions.



\begin{figure}
  \centering
  \begin{subfigure}[t]{0.33\columnwidth}
  \centering
  \includegraphics[height=3.3cm]{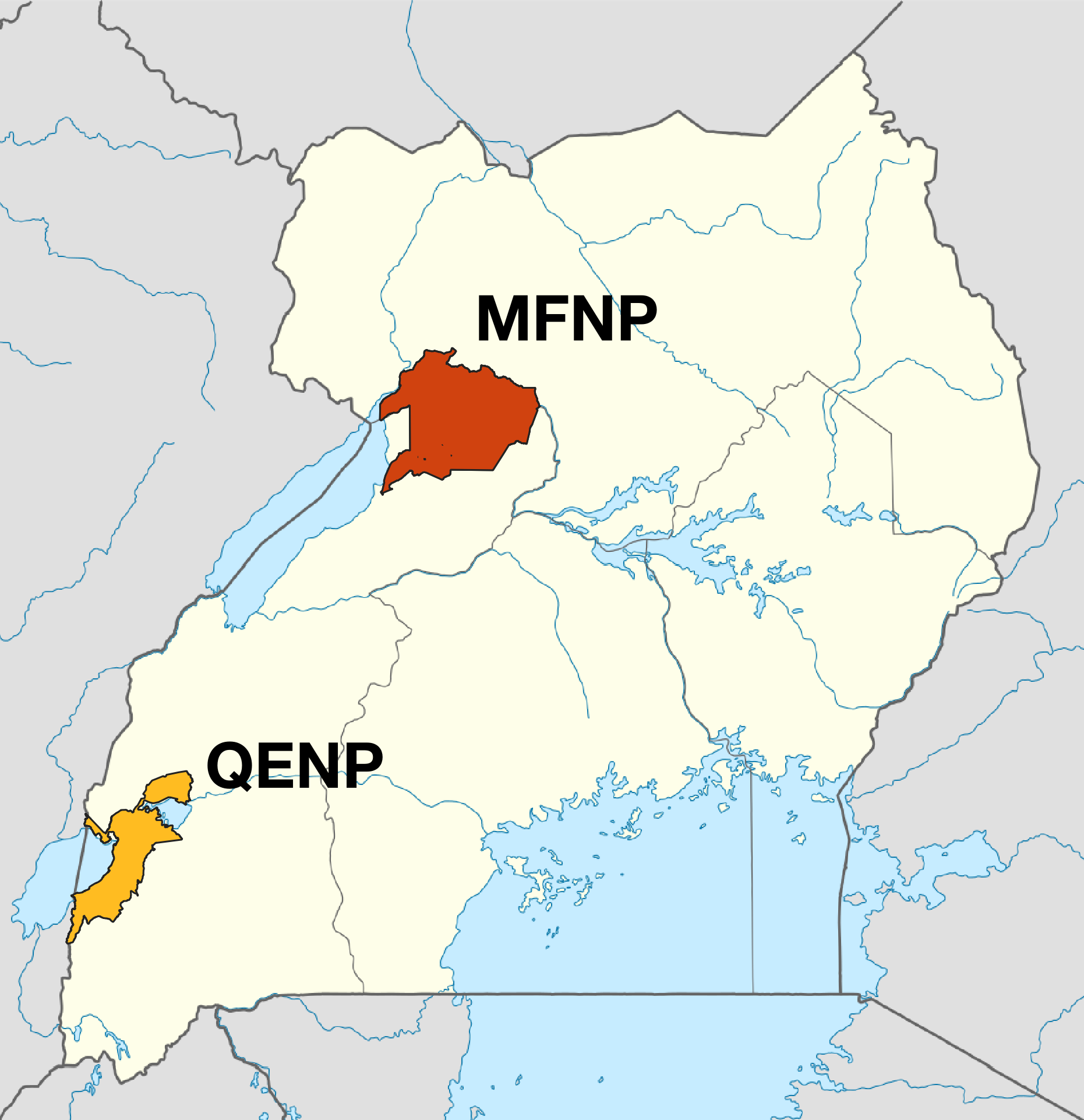}
  \end{subfigure}
  ~
  \begin{subfigure}[t]{0.63\columnwidth}
  \centering
  \includegraphics[height=3.3cm]{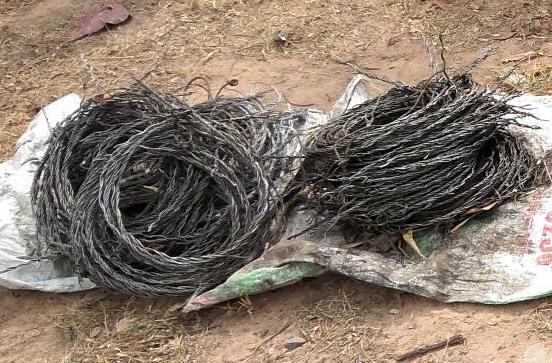}
  \end{subfigure}
  \caption{(Left) Map of Uganda, with locations of MFNP and QENP shown. (Right) Dozens of snares confiscated by rangers.}
  \label{fig:map-snares}
\end{figure}

%
%



\section{Background}



Researchers in both AI and conservation biology have independently pursued the question: do ranger patrols deter future poaching activity? Despite these efforts, most past work have been unable to find strong evidence of deterrence. Inconclusive results have come from trying to detect deterrence across an entire park or failing to properly account for the effect of patrol effort. This open question has important implications on a growing body of game theory work that aims to accurately model adversarial behavior. 


\paragraph{Past game-theoretic models.} A realistic model of deterrence is critical both to predict poaching and to design effective defender strategies. The AI literature has made assumptions of a deterrence effect in their machine learning and game theory models. In building predictive models, predictive models add past patrol effort as a covariate, leaving the machine learning algorithm to freely learn the impact without imposing any constraints \cite{gholami2018adversary,xu2020stay}. On the game theory side, Yang et al.\ \shortcite{yang2014adaptive} models the problem as a Stackelberg security game by assuming the poachers are reacting to the rangers' current action, rather than being deterred by past activity. Nguyen et al.\ \shortcite{nguyen2016capture} stipulates that poachers' action depends on the poacher behavior at the previous timestep, but not on ranger effort. Fang et al.\ \shortcite{fang2015security} provides an adversary model that responds to a convex combination of the defender strategy in recent rounds, which generalizes the deterrence effect.

\paragraph{Park-wide deterrence.} Despite a survey of Ugandan villagers which reveals that increased law enforcement would be most effective at deterring them from poaching \cite{harrison2015profiling}, conservation biologists have acknowledged that empirically showing deterrence remains a challenging task \cite{critchlow2017improving,dancer2019evaluation}. Dancer \shortcite{dancer2019evaluation} studies deterrence effects across four protected areas by analyzing change in patrol effort versus change in catch per unit effort (observations of illegal activity per kilometer patrolled), looking for deterrence across the entire park rather than within small regions of the park, as we do here. Dancer finds that increased patrols deter snare activity in only one of four sites, with a weak correlation at a one-month timescale. Dobson et al.\ \shortcite{dobson2019detecting} take a similar approach, but conduct analysis only on synthetic data which prevents us from drawing any conclusions from their work. 

\paragraph{Region-specific deterrence.} A question of particular importance for strategic patrol planning is whether deterrence can be detected at a finer scale, within a single park. The following literature investigates the impact of patrols on future likelihood of poaching at the $1 \times 1$~km or $500 \times 500$~m resolution. 
Plumptre et al.\ \shortcite{plumptre2014efficiently} find reduced levels of poaching activity within $8$~km of patrol posts. They conjecture that this spatial pattern demonstrates deterrence, due to the increased ranger presence in the area, but did not analyze whether poaching decreases due to increased past patrol effort.
Several others report no clear evidence of deterrence \cite{barichievy2017armed,okelly2018robust}.

Two other papers do suggest region-specific deterrence, but do not properly account for the impact of patrol effort. 
Ford \shortcite{ford2017real} looked at the probability of detecting poaching as rangers exerted high effort at a region in one month then low effort the following month, compared it against the probability of detecting poaching in regions where rangers went from low effort to high effort. Ford observed that poaching observations decreased in regions where rangers changed from high to low effort, but that decrease may be explained by the fact that by exerting low effort, rangers are less likely to find snares. Conversely, when observing more poaching in areas that went from low to high effort, that increase may simply be due to rangers being more thorough in finding snares. Thus, without accounting for the effect of current patrol effort, we cannot isolate the effect of past patrols. 
Moore et al.\ \shortcite{moore2018ranger} analyze poaching data from Nyungwe National Park in Rwanda to show that the probability that poaching abates in a cell increases in areas that rangers have visited more frequently. However, their analysis do not control for differences in current effort. The rangers may spend more time patrolling regions that they infrequently patrolled before and hence find more snares because of higher effort, which could explain the pattern without suggesting deterrence. 

One question that has been overlooked in past work is exploring where poachers are deterred \textit{to}. That is, are poachers deterred out of the park (thus producing a large-scale deterrence effect across the entire park), or are poachers deterred to other targets in the park? We provide evidence of the latter. 










\section{Domain}



The poaching data we use come from two national parks in Uganda, Queen Elizabeth and Murchison Falls. They are home to elephants, hippos, lions, and leopards alongside other mammals and more than 500 species of birds. Poachers hunt in these parks for both commercial (e.g., elephants) and noncommercial (e.g., bushmeat for personal consumption) wildlife \cite{critchlow2015spatiotemporal}. We use patrol data from 2010--2016.



We divide each park into $N$ targets, each of which is a $1 \times 1$~km region. We discretize the data both spatially, into the $N$ targets, and temporally. For the analysis in Section~\ref{sec:logistic-model}, we use three different temporal discretizations: one month, three month, and one year. For each target at each time step, we calculate the total ranger patrol effort (measured in kilometers patrolled) and count the number of instances of illegal activity detected. The patrol effort is constructed from 138.4k GPS waypoints in QENP and 94.7k in MFNP. Observations of illegal activity is predominantly in the form of snares, but can also include bullet cartridges, traditional weapons, or direct encounters with poachers. 

Each target is associated with a set of static geospatial features. These features include distance to park boundary, roads, permanent rivers, semi-permanent rivers, lakes, towns, villages, and patrol posts; slope, NPP, and wetness; and animal density estimates of Uganda kob, waterbuck, Jackson's hartebeest, topi, African buffalo, African elephant, warthog and giraffe.

\section{Analysis}

We want to understand the relationship between ranger actions in the past on poacher behavior in the present. To do so, we learn the impact of ranger actions in the previous timestep---first patrol effort in kilometers walked, then number of snare confiscations---on the number of instances of illegal activity detected in the current timestep. In doing so, we find clear evidence of deterrence in that higher levels of past patrols reduce the likelihood of poaching, even when accounting for current patrol effort. We also find that more intensive past patrols on neighboring targets \textit{increase} the likelihood of poaching, suggesting displacement. 


\begin{figure}
  \centering
  \begin{tikzpicture}


{
\small

\node[color=violet!40!black, fill=violet!15](feat) at (.5,4.25) {features};
\node[color=violet!40!black, fill=violet!15](attr) at (.5,3.25) {attractiveness};
\node[color=green!40!black, fill=green!10](peff) at (6,4) {past patrol effort};
\node[color=red!60!black, fill=red!10](pill) at (4.5,3) {past illegal activity};
\node[color=green!40!black, fill=green!10](ceff) at (3,2) {current patrol effort};
\node[color=red!60!black, fill=red!10](cill) at (3,1) {current illegal activity};
}

\draw[->](feat) -- (attr);
\draw[->](peff) -- (pill);
\draw[->](pill) -- (ceff);
\draw[->](ceff) -- (cill);
\draw[->](attr) -- (pill);
\draw[->](peff) to[out=-70,in=25] (cill);
\draw[->](attr) to[out=-90, in=155] (cill);




\end{tikzpicture}
  \caption{Relationship between the features, patrol effort, and illegal activity at each target. Attractiveness, current patrol effort, and past patrol effort directly impact our likelihood of detecting illegal activity in the current timestep.}
  \label{fig:logistic_model}
\end{figure}
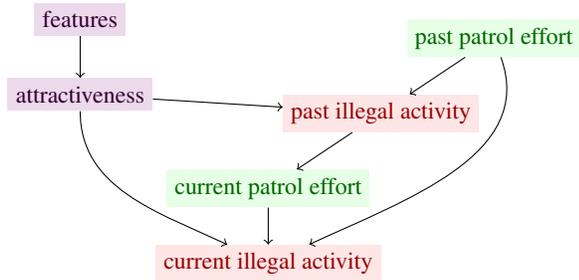

\subsection{Impact of past patrol effort}
\label{sec:logistic-model}

We learn a logistic model to understand the effect of current and past patrol effort on poaching activity. We know that poachers have an underlying preference for poaching spots based on convenience and their understanding of how likely a placed snare is to successfully trap an animal. Their preference can be understood as a measure of relative attractiveness between the targets, which is dependent on the geospatial features of each target. We only know about the poaching activity at targets that we visit. Furthermore, our likelihood of detecting poaching activity in each time period depends on the amount of effort that we exert. 
See Figure~\ref{fig:logistic_model} for a diagram of the relationship between each of these aspects. As shown, the factors that directly influence the probability of detecting illegal activity in the current timestep are attractiveness, current patrol effort, and past patrol effort. Our goal is to learn the relative influence of each of these factors. 

Each target in the park has its own attractiveness value, which is unknown. We let the attractiveness of each target be a parameter learned by the model, rather than trying to fit as a function of the features. Past work has shown that we can learn good models of poaching likelihood based on the features \cite{gholami2018adversary,xu2020stay}, but rely on complex models. A linear model would not be able to learn the interaction effects of different features, thus would learn a significantly inaccurate estimate of attractiveness. For the purpose of this analysis, we do not care about the individual attractiveness learned for each target, only their values relative to the impact of past patrol and current patrol effort. Let $a_i$ be the attractiveness of target $i$. We have the parameter $\beta$ as the coefficient on current patrol effort, which measures the detection likelihood, and $\gamma$ as the coefficient on past patrol effort, which measures the deterrence effect we are trying to isolate. Thus for a park with $N$ targets, we are learning $N+2$ total parameters. 

Specifically, we learn using the Adam optimizer \cite{kingma2015adam} the probability of detecting illegal activity in target $i$ as a linear combination of
\begin{align}
a_i + \beta \cdot \small{\texttt{curr\_effort}} + \gamma \cdot \small{\texttt{past\_effort}} \ ,
\end{align}
which is then squashed through the logistic function. 

See Table~\ref{tab:coef-effort} for the learned values of the average attractiveness of each target $\overline{a_i}$, the coefficient on current effort $\beta$, and the coefficient on past effort $\gamma$. The rows specify the timestep over which we study this effect. For example, year/3mo looks at the impact of a year of previous patrol effort on likelihood of detecting illegal activity in the subsequent three months. 
The inputs are normalized, so the coefficient can be interpreted as the effect of one standard deviation of effort on poaching probability, before being squashed by the logistic function. For example, $\beta = 0$ indicates the rangers are exerting a historically average amount of effort. To get a sense of the un-normalized results, we can analyze the standard deviation used to scale the original data. For example, for QENP 3 month, standard deviation of \texttt{curr\_effort} was $1.597$, and of \texttt{past\_effort} was $1.373$. 


\begin{table}
\caption {Learned coefficients, where $\gamma = $ past patrol effort} \label{tab:coef-effort} 
\begin{center}
\begin{tabular}{ r|ccc } 
\Xhline{2\arrayrulewidth}
& $\overline{a_i}$ & $\beta$ & $\gamma$ \\ 
\hline
& \multicolumn{3}{c}{MFNP} \\
\hline
3mo/3mo & $-9.284$ & $1.076$ & $-0.162$ \\ 
year/3mo & $-9.287$ & $1.062$ & $-0.216$ \\ 
year/year & $-8.56$3 & $2.159$ & $-0.306$ \\ 
\hline
& \multicolumn{3}{c}{QENP} \\
\hline
1mo/1mo & $-9.285$ & $1.074$ & $-0.165$ \\ 
3mo/3mo & $-10.624$ & $0.685$ & $-0.077$ \\ 
year/1mo & $-9.287$ & $1.061$ & $-0.217$ \\ 
year/3mo & $-10.629$ & $0.676$ & $-0.042$ \\ 
year/year & $-8.559$ & $2.159$ & $-0.306$ \\ 
\Xhline{2\arrayrulewidth}
\end{tabular}
\end{center}
\end{table}


The learned value of $\gamma$, the coefficient on past patrol effort, is negative across all datasets and settings. \textit{Thus, increased past patrol effort does have a measurable impact on deterring poaching.} Note that the attractiveness of each target is a relatively large negative number. This may seem counterintuitive, but $0$ in a logistic function corresponds with a $0.5$ probability, so negative values simply indicate the probability of detecting poaching is below $0.5$. In practice, observations of illegal activity are relatively rare, so we expect these values to be negative. The value $\overline{a_i}$ indicates the average base attractiveness value. The standard deviation of $a_i$ is between 4.34 and 4.98.

\subsection{Impact of past snare confiscation}

Just as rangers do not have perfect detection of poachers, poachers are also unable to perfectly observe the actions of rangers. In fact, poachers and rangers are unlikely to be at the same target at the same time. Out of 89,491 patrol observations over the period from 2010 to 2016 in QENP, 2,063 observations were of poaching activity, and only 111 of those were direct observations of poachers on the ground (54 of which led to arrests or fines, the remainder being unsuccessful pursuits). Based on the rarity of poachers and rangers being in direct contact, poachers are likely only to observe patrol effort when the rangers successfully confiscate a snare. Therefore, we hypothesized that detections of illegal activity---where rangers actually remove snares---would therefore have a greater impact on deterring poaching in the next time period. 

To test this hypothesis, we repeat our analysis from earlier, but replace past patrol effort with past detections of illegal activity. 
The relationship is therefore
\begin{align}
a_i + \beta \cdot \small{\texttt{curr\_effort}} + \rho \cdot \small{\texttt{past\_illegal}} \ .
\end{align}

\begin{table}
\caption {Learned coefficients, where $\rho = $ past illegal activity} \label{tab:coef-illegal} 
\begin{center}
\begin{tabular}{ r|ccc } 
\Xhline{2\arrayrulewidth}
& $\overline{a_i}$ & $\beta$ & $\rho$ \\ 
\hline
& \multicolumn{3}{c}{MFNP} \\
\hline
3mo/3mo & $-9.283$ & $1.066$ & $-0.134$ \\ 
year/3mo & $-9.314$ & $1.086$ & $-0.306$ \\ 
year/year & $-8.609$ & $2.290$ & $-0.517$ \\ 
\hline
& \multicolumn{3}{c}{QENP} \\
\hline
1mo/1mo & $-9.285$ & $1.066$ & $-0.135$ \\ 
3mo/3mo & $-10.632$ & $0.688$ & $-0.097$ \\ 
year/1mo & $-9.312$ & $1.085$ & $-0.307$ \\ 
year/3mo & $-10.647$ & $0.693$ & $-0.186$ \\ 
year/year & $-8.614$ & $2.291$ & $-0.516$ \\  
\Xhline{2\arrayrulewidth}
\end{tabular}
\end{center}
\end{table}

The learned coefficients are shown in Table~\ref{tab:coef-illegal}. Again, the coefficient $\rho$ is negative in all cases, indicating that increased detection of illegal activity deters future poaching. 
Furthermore, the values for $\rho$ are higher when using past illegal activity than when past effort. This result confirms our hypothesis that confiscating snares, which poachers will directly observe, has a greater impact in deterring poaching than simply walking more during a patrol. 

Observe that the effect is strongest when analyzing on a year-by-year basis. That is, increased patrolling sustained over an entire year has a stronger ability to deter poaching compared to increased patrol effort across a single month or a period of three months.

\subsection{Impact of patrolling nearby targets}
Ideally, when poachers are deterred by ranger patrols, they would leave the park completely and desist their hunt of wildlife. Alternatively, they may move to other areas of the park. We show that the latter appears to be true. 

We study the spatial relationship between neighboring targets to see whether increased patrolling in one region may deter poachers toward an adjacent region. To do so, we look at the cumulative past patrol effort of neighboring targets, using three spatial resolutions: $3 \times 3$, $5 \times 5$, and $7 \times 7$. Let \texttt{past\_neighbors} be the sum of the past instances of poaching activity on neighboring targets, so we learn
\begin{align}
a_i + \beta \cdot \small{\texttt{curr\_effort}} + \rho \cdot \small{\texttt{past\_illegal}} + \eta \cdot \small{\texttt{past\_neighbors}}
\end{align}
where $\eta$ is the coefficient on past poaching observations on neighboring cells. See Table~\ref{tab:coef-neighbors} for the learned coefficients, using data from QENP. All values of $\eta$ are positive, indicating that greater past detection of illegal activity on neighboring areas increases the likelihood of poaching on a target. This result is consistent across the three spatial resolutions, and strongest for the narrowest window of $3 \times 3$. Observe as well that the values for $\overline{a_i}$, $\beta$, and $\gamma$ are remarkably consistent, demonstrating the robustness of our findings. 


\begin{table}
\caption {Learned coefficients, with neighbors included} \label{tab:coef-neighbors} 
\begin{center}
\begin{tabular}{ r|cccc } 
\Xhline{2\arrayrulewidth}
& $\overline{a_i}$ & $\beta$ & $\rho$ & $\eta$ \\ 
\hline
$3 \times 3$ & $-10.627$ & $0.687$ & $-0.098$ & $0.399$ \\ 
$5 \times 5$ & $-10.634$ & $0.689$ & $-0.096$ & $0.383$ \\ 
$7 \times 7$ & $-10.632$ & $0.689$ & $-0.096$ & $0.562$ \\ 
\Xhline{2\arrayrulewidth}
\end{tabular}
\end{center}
\end{table}

\section{Discussion and conclusion}

Our results offer compelling evidence that ranger patrols are indeed effective at deterring poaching, substantiating the value of ranger efforts to wildlife conservation beyond the direct effect of removing snares. Additionally, rangers may be able to spot-patrol to deter poachers from a specific region, perhaps one that has exceptionally valuable animal habitat. 


Our finding that past observations of illegal activity has the most measurable deterrence effect gives us reason for optimism. Suppose to the contrary that only km of patrol effort influences deterrence. Accordingly, deterring poachers would require we hire more rangers to cover more ground. However, illegal activity implies that \textit{we can achieve significant deterrence with the same number of ranger resources} by focusing on increasing their effectiveness, specifically targeting areas that are at higher risk of having snares. Thus, the same number of rangers at the same cost can have a larger impact on preventing poaching. This effect is on top of the direct impact of removing more snares without increasing ranger resources. Future game theoretic algorithms should ideally optimize for this result. 

Furthermore, the displacement effect that we uncover provides better insight for modeling adversarial behavior and does not suggest that ranger patrols are futile in that they simply move poachers around the park. An open question for future work would be to more clearly explore the degree to which deterrence occurs across an entire park, in successfully pushing poachers out and preserving the biodiversity within these protected areas. 





\bibliographystyle{named}
\bibliography{short,ref}

\begin{thebibliography}{}

\bibitem[\protect\citeauthoryear{Barichievy \bgroup \em et al.\egroup
  }{2017}]{barichievy2017armed}
Chris Barichievy, Lawrence Munro, Geoffrey Clinning, Brendan Whittington-Jones,
  and Gavin Masterson.
\newblock Do armed field-rangers deter rhino poachers? an empirical analysis.
\newblock {\em Biological Conservation}, 209:554--560, 2017.

\bibitem[\protect\citeauthoryear{Critchlow \bgroup \em et al.\egroup
  }{2015}]{critchlow2015spatiotemporal}
Rob Critchlow, Andrew~J Plumptre, Margaret Driciru, Aggrey Rwetsiba, Emma~J
  Stokes, Charles Tumwesigye, Fred Wanyama, and CM~Beale.
\newblock Spatiotemporal trends of illegal activities from ranger-collected
  data in a ugandan national park.
\newblock {\em Conservation Biology}, 29(5):1458--1470, 2015.

\bibitem[\protect\citeauthoryear{Critchlow \bgroup \em et al.\egroup
  }{2017}]{critchlow2017improving}
Rob Critchlow, Andrew~J Plumptre, Bazil Alidria, Mustapha Nsubuga, Margaret
  Driciru, Aggrey Rwetsiba, F~Wanyama, and Colin~M Beale.
\newblock Improving law-enforcement effectiveness and efficiency in protected
  areas using ranger-collected monitoring data.
\newblock {\em Conservation Letters}, 10(5):572--580, 2017.

\bibitem[\protect\citeauthoryear{Dancer}{2019}]{dancer2019evaluation}
Anthony Dancer.
\newblock {\em On the evaluation, monitoring and management of law enforcement
  patrols in protected areas}.
\newblock PhD thesis, University College London, 2019.

\bibitem[\protect\citeauthoryear{Dobson \bgroup \em et al.\egroup
  }{2019}]{dobson2019detecting}
Andrew~DM Dobson, EJ~Milner-Gulland, Colin~M Beale, Harriet Ibbett, and Aidan
  Keane.
\newblock Detecting deterrence from patrol data.
\newblock {\em Conservation Biology}, 33(3):665--675, 2019.

\bibitem[\protect\citeauthoryear{Fang \bgroup \em et al.\egroup
  }{2015}]{fang2015security}
Fei Fang, Peter Stone, and Milind Tambe.
\newblock When security games go green: Designing defender strategies to
  prevent poaching and illegal fishing.
\newblock In {\em Proc. of IJCAI-15}, 2015.

\bibitem[\protect\citeauthoryear{Ford}{2017}]{ford2017real}
Benjamin~John Ford.
\newblock {\em Real-world evaluation and deployment of wildlife crime
  prediction models}.
\newblock PhD thesis, University of Southern California, 2017.

\bibitem[\protect\citeauthoryear{Gholami \bgroup \em et al.\egroup
  }{2018}]{gholami2018adversary}
Shahrzad Gholami, Sara Mc~Carthy, Bistra Dilkina, Andrew Plumptre, Milind
  Tambe, Margaret Driciru, Fred Wanyama, Aggrey Rwetsiba, Mustapha Nsubaga,
  Joshua Mabonga, et~al.
\newblock Adversary models account for imperfect crime data: Forecasting and
  planning against real-world poachers.
\newblock In {\em Proc. of AAMAS-18}, pages 823--831, 2018.

\bibitem[\protect\citeauthoryear{Harrison \bgroup \em et al.\egroup
  }{2015}]{harrison2015profiling}
Mariel Harrison, Julia Baker, Medard Twinamatsiko, and EJ~Milner-Gulland.
\newblock Profiling unauthorized natural resource users for better targeting of
  conservation interventions.
\newblock {\em Conservation Biology}, 29(6):1636--1646, 2015.

\bibitem[\protect\citeauthoryear{Kingma and Ba}{2015}]{kingma2015adam}
Diederik~P Kingma and Jimmy Ba.
\newblock Adam: A method for stochastic optimization.
\newblock In {\em Proc. of ICLR-15}, 2015.

\bibitem[\protect\citeauthoryear{Levitt}{1998}]{levitt1998increased}
Steven~D Levitt.
\newblock Why do increased arrest rates appear to reduce crime: deterrence,
  incapacitation, or measurement error?
\newblock {\em Economic inquiry}, 36(3):353--372, 1998.

\bibitem[\protect\citeauthoryear{Moore \bgroup \em et al.\egroup
  }{2018}]{moore2018ranger}
Jennifer~F Moore, Felix Mulindahabi, Michel~K Masozera, James~D Nichols,
  James~E Hines, Ezechiel Turikunkiko, and Madan~K Oli.
\newblock Are ranger patrols effective in reducing poaching-related threats
  within protected areas?
\newblock {\em Journal of Applied Ecology}, 55(1):99--107, 2018.

\bibitem[\protect\citeauthoryear{Nguyen \bgroup \em et al.\egroup
  }{2016}]{nguyen2016capture}
Thanh~H Nguyen, Arunesh Sinha, Shahrzad Gholami, Andrew Plumptre, Lucas Joppa,
  Milind Tambe, Margaret Driciru, Fred Wanyama, Aggrey Rwetsiba, Rob Critchlow,
  et~al.
\newblock Capture: A new predictive anti-poaching tool for wildlife protection.
\newblock In {\em Proc. of AAMAS-16}, pages 767--775, 2016.

\bibitem[\protect\citeauthoryear{O'Kelly \bgroup \em et al.\egroup
  }{2018}]{okelly2018robust}
Hannah~J O'Kelly, J~Marcus Rowcliffe, Sarah~M Durant, and EJ~Milner-Gulland.
\newblock Robust estimation of snare prevalence within a tropical forest
  context using n-mixture models.
\newblock {\em Biological Conservation}, 217:75--82, 2018.

\bibitem[\protect\citeauthoryear{Plumptre \bgroup \em et al.\egroup
  }{2014}]{plumptre2014efficiently}
Andrew~J Plumptre, Richard~A Fuller, Aggrey Rwetsiba, Fredrick Wanyama, Deo
  Kujirakwinja, Margaret Driciru, Grace Nangendo, James~EM Watson, and Hugh~P
  Possingham.
\newblock Efficiently targeting resources to deter illegal activities in
  protected areas.
\newblock {\em Journal of Applied Ecology}, 51(3):714--725, 2014.

\bibitem[\protect\citeauthoryear{Xu \bgroup \em et al.\egroup
  }{2020}]{xu2020stay}
Lily Xu, Shahrzad Gholami, Sara~Mc Carthy, Bistra Dilkina, Andrew Plumptre,
  Milind Tambe, Rohit Singh, Mustapha Nsubuga, Joshua Mabonga, Margaret
  Driciru, et~al.
\newblock Stay ahead of poachers: Illegal wildlife poaching prediction and
  patrol planning under uncertainty with field test evaluations.
\newblock In {\em Proc. of ICDE-20}, 2020.

\bibitem[\protect\citeauthoryear{Yang \bgroup \em et al.\egroup
  }{2014}]{yang2014adaptive}
Rong Yang, Benjamin Ford, Milind Tambe, and Andrew Lemieux.
\newblock Adaptive resource allocation for wildlife protection against illegal
  poachers.
\newblock In {\em Proc. of AAMAS-14}, pages 453--460. Citeseer, 2014.

\bibitem[\protect\citeauthoryear{‘t Sas-Rolfes \bgroup \em et al.\egroup
  }{2019}]{gulland2019illegal}
Michael ‘t Sas-Rolfes, Daniel~WS Challender, Amy Hinsley, Diogo
  Ver{\'\i}ssimo, and EJ~Milner-Gulland.
\newblock Illegal wildlife trade: Scale, processes, and governance.
\newblock {\em Annual Review of Environment and Resources}, 44:201--228, 2019.

\end{thebibliography}


\end{document}